\documentclass[12pt]{article}
\usepackage[cp1251]{inputenc}
\usepackage[russian]{babel}
\usepackage{hyperref,amsfonts,amssymb,theorem}

\def\bq{ \begin{equation} }
\def\eq{ \end{equation} }
\def\ben{ \begin{eqnarray} }
\def\en{ \end{eqnarray} }

\theorembodyfont{\sl}

\theorembodyfont{\rm}

\def\Ac{\textsf{S}}
\def\dsize{\displaystyle}

\def\p{ \partial }

\def\bq{ \begin{equation} }
\def\eq{ \end{equation} }
\def\ben{ \begin{eqnarray} }
\def\en{ \end{eqnarray} }

\def\frac#1#2{{#1\over #2}}

\def\on#1#2{\mathop{\vbox{\ialign{##\crcr\noalign{\kern2pt}
$\scriptstyle{#2}$\crcr\noalign{\kern2pt\nointerlineskip}
\kern-2pt$\hfil\displaystyle{#1}\hfil$\crcr}}}\limits}

\newcommand{\be}{\begin{equation}}
\newcommand{\ee}{\end{equation}}
\newcommand{\dx}{\partial _{q_{1}}}
\newcommand{\dy}{\partial _{q_{2}}}

\newcommand{\bear}{\begin{array}}
\newcommand{\eear}{\end{array}}

\def\ba{\begin{array}}
\def\ea{\end{array}}

\begin{document}

\vspace{1cm} \centerline {\LARGE \textbf{Pairs of commuting Hamiltonians,}} \vspace{0.3cm} \centerline {\LARGE
\textbf{quadratic in momenta}} \vskip1cm \hfill
\begin{minipage}{13.5cm}
\baselineskip=15pt {\bf
 V.G. Marikhin ${}^{1}$ and
 V.V. Sokolov ${}^{1}$} \\ [2ex]
{\footnotesize
${}^{1}$ L.D. Landau Institute for Theoretical Physics RAS, Moscow, Russia}\\

In the case of two degree system the pairs of quadratic in momenta Hamiltonians
commuting according the standard Poisson bracket are considered.
The new many-parametrical families of such pairs are founded.
The universal method of constructing the full solution of Hamilton - Jacobi equation in terms of integrals
on some algebraic curve is proposed. For some examples this curve is non-hyperelliptic covering over
the elliptic curve.

\end{minipage}

\vglue1cm \textbf{Address}:  \qquad  Landau Institute for Theoretical Physics RAS, Kosygina st.2 ,Moscow, Russia,  119334,

\textbf{E-mail}: \qquad mvg@itp.ac.ru, \,  sokolov@itp.ac.ru
\newpage


\section{Pairs of quadratic hamiltonians}
In papers  \cite{winter1,ferap,ferapfor,winter2,yehia, marsok,marsok2} the problem
of the commuting pairs of Hamiltonians quadratic in momenta was considered.
\setcounter{equation}{0}

Consider pair of Hamiltonians in the form

\be \label{quadr}
H=a p_{1}^{2}+ 2b p_1p_2+c p_{2}^{2}+d p_{1}+e p_{2}+f,
\ee
\be \label{quadr1}
K=A p_{1}^{2}+ 2B p_1p_2+C p_{2}^{2}+D p_{1}+E p_{2}+F,
\ee
commuting with respect to standart poisson bracket
$\lbrace p_{\alpha},q_{\beta}\rbrace=\delta_{\alpha
\beta}$. The coefficients in formulas (\ref{quadr}),(\ref{quadr1}) -
some (locally) analitical functions of the variables
$q_{1},q_{2}$.

{\bf Theorem 1.} {\it Any pairs of commuting Hamiltonians}
(\ref{quadr})-(\ref{quadr1})  {\it can be canonically transformed by}
$$
\hat P_{1}= P_{1}+\frac{\p F(s_{1}, s_{2})}{\p s_{1}}, \qquad
\hat  P_{2}= P_{2}+\frac{\p F(s_{1}, s_{2})}{\p sq_{2}}
$$
{\it to the pair of the form}
\be\label{HKG} H=\frac{U_1-U_2}{s_1-s_2},\qquad
K=\frac{s_2\,U_1-s_1\,U_2}{s_1-s_2}, \ee
{\it where}
\be
\begin{array}{l} \label{Lxyg}
U_1=\dsize{S_{1}(s_1)P_1^2+\frac{\sqrt{S_{1}(s_1)S_{2}(s_2)}Z_{s_1}}{(s_1-s_2)}P_2-
\frac{S_{1}(s_1)Z_{s_1}^2}{4(s_1-s_2)^2}
+V_1(s_1,s_2),}\\[8mm]
U_2=\dsize{S_{2}(s_2)P_2^2-\frac{\sqrt{S_{1}(s_1)S_{2}(s_2)}Z_{s_2}}
{(s_1-s_2)}P_1-\frac{S_{2}(s_2)Z_{s_2}^2}{4(s_2-s_1)^2}
+V_2(s_1,s_2),}
\end{array}
\ee
\be\begin{array}{l}\label{tails}
V_1=\dsize{\frac{1}{2}\sqrt{S_{1}(s_1)}\,\dx\left(\sqrt{S_{1}(s_1)})
\frac{Z_{s_1}^2}{s_1-s_2}\right)+f_{1}(s_1)},\\[5mm]
V_2=\dsize{\frac{1}{2}\sqrt{S_{2}(s_2)}\,\dy\left(\sqrt{S_{2}(s_2)})
\frac{Z_{s_2}^2}{s_2-s_1}\right)+f_{2}(s_2)}
\end{array}\ee
{\it for some functions} $Z(s_1,s_2)$, $S_{i}(s_{i})$ {\it and}
$f_{i}(s_{i})$. {\it Poisson bracket} $\lbrace H,K\rbrace$ {\it
equals to zero if and only if} 
\be\label{usl1}
Z_{s_1,s_2}=\frac{Z_{s_1}-Z_{s_2}}{2(s_2-s_1)} \ee {\it and} 
\be\label{usl2}
 \left( Z_{s_1} \frac{\partial}{\partial s_2}- Z_{s_2} \frac{\partial}{\partial s_1}\right)
 \, \left( \frac{V_1-V_2}{s_1-s_2}\right)=0.
\ee
{\bf Proof}
We introduce new coordinates $s_1,s_2$, such that the quadratic parts of $H,K$  (\ref{quadr},\ref{quadr1}) are diagonal:
Let  $s_1,s_2$ be the roots of equations
\be\label{eg_s}
\Phi(s,q_1,q_2)=(B-bs)^2-(A-as)(C-cs)=0,
\ee

Then the canonical transformation 
\be\label{can}
(q_1,q_2,p_1,p_2)\rightarrow (s_1,s_2,P_1,P_2): p_1=-(\frac{\Phi^1_{q_1}}{\Phi^1_{s_1}}P_1+\frac{\Phi^2_{q_1}}{\Phi^2_{s_2}}P_2),\quad
p_2=-(\frac{\Phi^1_{q_2}}{\Phi^1_{s_1}}P_1+\frac{\Phi^2_{q_2}}{\Phi^2_{s_2}}P_2),
\ee
where $\Phi^i=\Phi(s_i,q_1,q_2)$ under conditions  $\lbrace H,K\rbrace=0$ transforms pairs
(\ref{quadr}),(\ref{quadr1}) to the form
\be\label{HKG2} H=\frac{U_1-U_2}{s_1-s_2},\qquad
K=\frac{s_2\,U_1-s_1\,U_2}{s_1-s_2}, \ee
where
\be\label{quadr_can}
U_1=S_1(s_1)P_1^2+\tilde{d}P_1+\tilde{e}P_2+\tilde{f},\quad U_2=S_2(s_1)P_2^2+\tilde{D}P_1+\tilde{E}P_2+\tilde{F},
\ee

where 
\be\label{S}
S_i(s_i)=\frac{1}{(\Phi^i_{q_i})^2}((as_i-A)(\Phi^i_{q_1})^2+2(bs_i-B)\Phi^i_{q_1}\Phi^i_{q_2}+(cs_i-C)(\Phi^i_{q_2})^2)
\ee

We calculate a Poisson bracket between H and K. Then the coefficient of $P_1^2, P_2^2, P_1P_2$ equal to  zero
iff 
$$\tilde{d}=2S_1(s_1)\frac{\partial F(s_1,s_2)}{\partial s_1},\;
\tilde{e}=\frac{\sqrt{S_{1}(s_1)S_{2}(s_2)}Z_{s_1}}{(s_1-s_2)},\;
\tilde{D}=-\frac{\sqrt{S_{1}(s_1)S_{2}(s_2)}Z_{s_2}}{(s_1-s_2)},\;
\tilde{E}=2S_2(s_2)\frac{\partial F(s_1,s_2)}{\partial s_2}
$$
where $Z(s_1,s_2),\;F(s_1,s_2)$ - some functions.
and 
\be\label{usl1_2}
Z_{s_1,s_2}=\frac{Z_{s_1}-Z_{s_2}}{2(s_2-s_1)} \ee
We apply the canonical transformation 
$$
\hat P_{1}= P_{1}+\frac{\p F(s_{1}, s_{2})}{\p s_{1}}, \qquad
\hat  P_{2}= P_{2}+\frac{\p F(s_{1}, s_{2})}{\p s_{2}}
$$
to equate $\tilde{d},\;\tilde{E}$ to zero.
Then the coefficient of $P_1, P_2$ equal to zero iff $U_1,U_2$ have the form as in formulation of Theorem 1. And finally the free coefficient in Poisson bracket equals to zero iff the equation (\ref{usl2}) of the Theorem 1 
is fulfilled just as expected.

The general analytical solution of Euler - Darboux equation
(\ref{usl1}) has near the line of singularities $x=y$ the following expansion:
$$
Z(x,y)=A+\hbox{ln}(x-y)\, B,\qquad  A=\sum_{0}^{\infty} a_{i}(x+y)\, (x-y)^{2
i}, \qquad
B=\sum_{0}^{\infty} b_{i}(x+y)\, (x-y)^{2 i},
$$
where $a_{0}$ and $a_{1}$ - some function. The other coefficients can be expressed
by these two functions and their derivatives. For example, $b_{0}=\frac{1}{2} a_{0}''$.

We insert this expantion into (\ref{usl2}) to obtain $B=0$. It is easy to check that any solution of the equation (\ref{usl1}) with $B=0$
has the form 
\be\label{genDE1} Z(x,y)=z_0+\delta
(x+y) +
(x-y)^2\sum\limits_{k=0}^{\infty}\frac{g^{(2k)}(x+y)}{2^{(2k)}k!(k+1)!}(x-y)^{2k},
\ee where $g(x)$ - some function and $z_{0}, \delta$ -
some constants. We call the function $g(x)$  {\it
as generating function} for (\ref{genDE1}). Without the loss of generality we choose  $z_{0}=0.$  The parameter $\delta$, is very important for classification of hamiltonians from Theorem 1. 

We find all the functions $Z,$ corresponding the rational generating functions $g$. Choosing $g(x)=x^{n}$,
we obtain the infinite set of polynomial solutions
$Z^{(n)}$ for (\ref{usl1}). In particular
$$
g(x)=1\iff Z^{(0)}(x,y)=(x-y)^2
$$
$$
g(x)=x\iff Z^{(1)}(x,y)=(x+y)(x-y)^2,$$
$$
g(x)=x^2\iff
Z^{(2)}(x,y)=\frac{1}{4}\left((x-y)^2+4(x+y)^2\right)(x-y)^2.$$

All set can be obtained by using 'creating' operator
$$ x^{2} \frac{\partial}{\partial x}+ y^{2}
\frac{\partial}{\partial y}-\frac{1}{2} (x+y), $$ acting on
$Z^{(0)}.$ The rational functions $g(x)=(x-\mu)^{-n}$ create one more class of exact solution
of equation (\ref{usl1}). For example
$$
g_{\mu}(x)=\frac{1}{4}\frac{1}{x-2 \mu}\iff
Z_{\mu}(x,y)=\sqrt{(\mu-x)(\mu-y)}+\frac{1}{2}(x+y)-\mu.
$$
The solution corresponding the poles of order
 $n\ge 2,$ can be obtained by differentiating the last formula by parameter $\mu.$
Because function $Z$ is linear by $g$ we obtained the solution 
$Z$ with rational generating function $g(x)=\sum_{i} c_{i} x^{i}+\sum_{i,j} d_{ij}
(x-\mu_{i})^{-j}. $

{\bf Hypothesis 1.} For all Hamiltonians
(\ref{HKG})-(\ref{usl2}) generating function $g$ is rational and has the form
 $g(x)=\frac{P(x)}{S(x)},$ where $P$ ш $S$ - some polynomials with $\hbox{deg} P<5,\, \hbox{deg} S<6.$

In papers \cite{yehia, marsok} the following solution of the system (\ref{usl1}), (\ref{usl2}) was considered:
$$
\begin{array}{c}
\displaystyle Z(x, y)=x+y, \qquad S_{1}(x)=S_{2}(x)=\sum_{i=0}^{6}
c_{i} x^{i}, \quad
\\[4mm]
\displaystyle f_{1}(x)=f_{2}(x)=-\frac{3}{4} c_{6}
x^{4}-\frac{1}{2} c_{5} x^{3}+\sum_{i=0}^{2} k_{i} x^{i},
\end{array}
$$
where $c_{i}, k_{i}$ - some constants. A very important fact is that Clebsch top and $so(4)$-Schottky-Manakov top \cite{shot, man, clebsch}
are the particular cases of this model \cite{marsok}. In paper
\cite{marsok} a full solution of Hamilton - Jacobi  equation of this model was obtained
in the form of some kind of separation of variables on a non-hyperelliptic curve of genus 4.

\section{Universal solution of Hamilton-Jacobi equation}

Let $H$ and $K$ have the form (\ref{HKG})-(\ref{tails}). Consider system $H=e_1,\, K=e_2$, where $e_i$ - some constants. Let
$p_1=F_1(x,y),\, p_2=F_2(x,y)$ - be its solution. We use short notation
  $x$ ш $y$ corresponding $q_1$ ш $q_2$. Jacobi's lemma gives
that if $\{H,\,K\}=0$, then $\frac{\p F_1}{\p
y}=\frac{\p F_2}{\p x}$. To find an action
$\Ac(x,y,e_1,e_2),$ it is enough to solve the following system
$$
\frac{\p}{\p x}\Ac=F_1, \qquad \frac{\p}{\p y}\Ac=F_2.
$$

We rewrite the system $H=e_1,\, K=e_2$ in the form
\be\label{aabb} p_1^2+a
p_2+b=0,\qquad p_2^2+A p_1+B=0, \ee
where
$$
a=\frac{Z_x}{x-y}\sqrt{\frac{S_2(y)}{S_1(x)}} ,\qquad
A=-\frac{Z_y}{x-y}\sqrt{\frac{S_1(x)}{S_2(y)}}
$$
$$b=-\frac{Z_x^2}{4(x-y)^2}+\frac{V_1-e_1x+e_2}{S_1(x)},\qquad
B=-\frac{Z_y^2}{4(x-y)^2}+\frac{V_2-e_1y+e_2}{S_2(y)}.
$$
It easy to find that \be\label{to12} 2 b_y+ A a_x+2 a A_x=0,\quad
2A a_y+a A_y+2 B_x=0. \ee
Using  (\ref{usl1}) and
(\ref{usl2}), it is easy to obtain the following identity
\be\label{to3} A b_x-a B_y+2A_x b-2a_y B=0. \ee

Using a standard technique of Lagrange resolvents (see f.e. \cite{prsol}), 
we rewrite system (\ref{aabb}) to a system 
\be\label{curuv1} u
v=\frac{1}{4}a A, \ee 
\be\label{curuv2} A u^3+4\frac{b}{a}u^2
v-4\frac{B}{A} u v^2-a v^3=0,\ee that is equivalent to the qubic equation on
 $u^2$. Let $(u_k, v_k),\, k=1,2,3$
be the solutions of  (\ref{curuv1}), (\ref{curuv2}) such that
$$
\begin{array}{c}
u_1^2+u_2^2+u_3^2=-b,\qquad v_1^2+v_2^2+v_3^2=-B\\[3mm]
\dsize u_1 u_2 u_3=-\frac{1}{8}a^2 A,\qquad v_1 v_2
v_3=-\frac{1}{8}A^2 a.
\end{array}
$$
Then, formulas
$$
\begin{array}{l} \label{u123v123}
p_1=u_1+u_2+u_3,\qquad p_2=v_1+v_2+v_3;\\
p_1=u_3-u_1-u_2,\qquad p_2=v_3-v_1-v_2;\\
p_1=u_2-u_1-u_3,\qquad p_2=v_2-v_1-v_3;\\
p_1=u_1-u_2-u_3,\qquad p_2=v_1-v_2-v_3\\
\end{array}
$$
define four solutions of (\ref{aabb}). Consider the first of them.

{\bf Lemma 1.} {\it For} $i=1,2,3$ {\it following equations are fullfiled}
$\frac{\p u_i}{\p y}=\frac{\p v_i}{\p x}.$

{\bf Prove.} Differentiating equations (\ref{curuv1}) and
(\ref{curuv2}) on $x$ and $y,$ we find $u_y$ and $v_x$ as the functions on
$u$ and $v.$ Then expressing $v$ through $u,$ we obtain that
$u_y=v_x$ is equivalent to identities  (\ref{to12}) and
(\ref{to3}). $\blacksquare$

Lemma 1 means, that in variables $u_1, u_2, u_3$ we find
''particular'' separation variables. Really $\Ac=\Ac_1+\Ac_2+\Ac_3$, where
$\Ac$ is the action, and functions $\Ac_i$ defined from a system
$$
\frac{\p}{\p x}\Ac_i=u_i, \qquad \frac{\p}{\p y}\Ac_i=v_i.
$$

Let's
$$
\label{uvdef}
u=\frac{1}{2}\frac{Z_x}{x-y}\sqrt{\frac{y-\xi}{x-\xi}},\qquad
v=-\frac{1}{2}\frac{Z_y}{x-y}\sqrt{\frac{x-\xi}{y-\xi}}\,.
$$
It easy to see that pair $(u,v)$ for all $\xi$ are a solution of (\ref{curuv1}). 
If $Z$ is a solution of  (\ref{usl1}), then $\frac{\p u}{\p
y}=\frac{\p v}{\p x}.$ Using this fact we introduce a function
$\sigma(x,y,\xi)$ so that
$$
\frac{\p \sigma}{\p x}=u, \qquad \frac{\p \sigma}{\p y}=v.
$$
In a case of rational function $g,$ corresponding function $Z$ 
is expressed through quadratic radicals and the function $\sigma$
can be obtained. Let's $\dsize Y=\frac{\p
\sigma}{\p \xi}$.

After multiplication of expression (\ref{curuv2}) by expression
$$-2\frac{\sqrt{S_1(x)}\sqrt{S_2(y)}\sqrt{x-\xi} \sqrt{y-\xi}\,(x-y)}{Z_x Z_y},$$
left side of  (\ref{curuv2}) can be written in the form
\be\label{curxyxi}
-e_2+e_1\xi+\frac{y-\xi}{x-y}\Big(V_1-\frac{S_1(x)Z_x^2}{4(x-\xi)(x-y)}\Big)-
\frac{x-\xi}{x-y}\Big(V_2+\frac{S_2(y)Z_y^2}{4(y-\xi)(x-y)}\Big).
\ee

{\bf Proposition 1}. {\it Let the expression (\ref{usl1}),
(\ref{usl2}) be fulfilled } . {\it Then the expression} (\ref{curxyxi}) {\it is a function of } $Y$ {\it and} $\xi$
variables only.

{\bf Prove}. We assign the function (\ref{curxyxi}) as
$\Psi(x,y,\xi)$. Consider Jacobian
$$
J=\frac{\p \Psi}{\p x} \frac{\p Y}{\p y}-\frac{\p \Psi}{\p y}
\frac{\p Y}{\p x}.
$$
We change $\frac{\p Y}{\p y}$ and $ \frac{\p Y}{\p x}$ to $\frac{\p
v}{\p \xi}$ and $ \frac{\p u}{\p \xi},$ respectevily, then 
Jacobian $J$ equals to zero identically  taking into account (\ref{usl1}), (\ref{usl2}). $\blacksquare$

Due to Proposition 1, the relation $\Psi(x,y,\xi)=0$ can be rewritten in the form $\phi(\xi,Y)=0.$ 
One can find the function  $\phi$ by assuming $y=x$.

Equation $\phi(\xi,Y)=0$ defines a curve, and the differentials of this curve define
the function of action  $S.$

We note $\xi_k(x,y)$,  where $k=1,2,3$, the roots of cubic equation $\Psi(x,y,\xi)=0.$

{\bf Theorem 2.} {\it The function of action} $\Ac$ {\it has the form}
\be\label{act} \Ac(x,y)=\sum\limits_{k=1}^3\Big(\,
\sigma(x,y,\xi_k)-\int\limits^{\xi_k} Y(\xi)\, d\xi\,\Big),  \ee
{\it where} $Y(\xi)$ - {\it alebraic function on the curve}
$\phi(\xi,Y)=0.$

{\bf Prove.}  We obtain $$\frac{\p }{\p
x}\Ac(x,y)=\sum\limits_{k=1}^3
\sigma_x(x,y,\xi_k)+\sum\limits_{k=1}^3\lbrace\,\sigma_{\xi}(x,y,\xi_k)-Y(\xi_k)\rbrace\xi_{k,x}=
\sum\limits_{k=1}^3u_k=p_1.$$ Analogously 
$$\frac{\p}{\p
y}\Ac(x,y)=p_2. \qquad \blacksquare$$

\section{Case of cubics}
Consider a case when the curve(\ref{curxyxi}) can be written in the form
$\tilde{\phi}(\xi,\eta)=0\Leftrightarrow \phi(\xi,Y),$ so that points $(\xi_1,\eta_1),(\xi_2,\eta_2),(\xi_3,\eta_3)$ 
lie on a straight line, that equivalent to definition 
$\eta=\xi a(x,y)+b(x,y),$ its substitution into $\tilde{\phi},$ 
gives a curve $\Psi(x,y,\xi)=0.$

Formula (\ref{curxyxi}) gives a curve in a new variables $\xi,\eta$  
 $-e_2+e_1\xi+\frac{C_2(\xi,\eta)}{C_1(\xi,\eta)}=0,$ where $C_1(\xi,\eta)\rightarrow 0$ at $x\rightarrow 0$ or
$y\rightarrow 0.$

Using reversible curve equation $C_1(\xi,\eta)=0\rightarrow \eta=f(\xi)$ using $\eta,$ we find the expressions for $a(x,y),\; b(x,y)$
$$
a(x,y)=\frac{f(x)-f(y)}{x-y},\; b(x,y)=\frac{y f(x)-x f(y)}{x-y}
$$ 

On the other hand the equivalence of the curve $\phi(\xi,Y)=0$ ш $\tilde{\phi}(\xi,\eta)=0$ 
gives 
$$Y_x\eta_y=Y_y\eta_x \Leftrightarrow u_{\xi}\eta_y=v_{\xi}\eta_x \Leftrightarrow
(\xi-y)Z_x\eta_y=(x-\xi)Z_y\eta_x,
$$
or $Z=Z(a),\quad b_x=-y\,a_x,\; b_y=-x\,a_y.$

\section{Examples}

In this Section we consider all the pairs of Hamiltonians known at the moment (\ref{HKG})-(\ref{usl2}).

\subsection{Class 1}

For the models of this class
\begin{equation} \label{sym}
S_1=S_2=S, \qquad f_1=f_2=f.
\end{equation}

{\bf Theorem 3.} {\it Let}
$$
g=\frac{\tilde G}{S},
\qquad \quad \tilde G=G-\frac{\delta}{10} S', \qquad \quad
 f=-\frac{4 \tilde
G^{2}}{S}-\frac{4 \delta}{3} \tilde G'-\frac{\delta^{2}}{12} S'',
$$
{\it  where}
$$
S(x)=s_{5} x^{5}+s_{4} x^{4}+s_{3} x^{3}+s_{2} x^{2}+s_{1}
x+s_{0}, \qquad G(x)=g_{3} x^{3}+g_{2} x^{2}+g_{1} x+g_{0},
$$
{\it where} $s_i, g_i, \delta$ - {\it  some constants. Then
functions} $S, \,f$ {\it  and function} $Z,$ {\it corresponding} (see.
\S 1) {\it generation function} $g,$ {\it  fulfill the systems}
(\ref{usl1}), (\ref{usl2}).

{\bf Remark.} Parameter $\delta$ from Theorem 3 coinsides
with parameter $\delta$ from (\ref{genDE1}). Consider the case $\delta=0$
in the formula (\ref{genDE1}),  Then all pairs of Hamiltonians (\ref{HKG})-(\ref{usl2}), (\ref{sym}),
that fullfil this condition are described by Theorem 3.

Consider a general case $$
S(x)=s_5(x-\mu_1)(x-\mu_2)(x-\mu_3)(x-\mu_4)(x-\mu_5),
$$
where $s_5\ne 0$ and all roots $\mu_i$ of polynomial $S$ are distinct.
then the function  $Z$ has the form
 \be\label{Z5roots} Z(x,\,y)=\sum\limits_{i=1}^5
\nu_i\sqrt{(\mu_i-x)(\mu_i-y)}, \ee where $\nu_{i}$ - some
constants.  Coeffitients $g_i$ and $\delta $ are expressed through constants 
$\nu_j$ from (\ref{genDE1}). For example, $2
\delta=-\sum \nu_{i}.$ Function $f$ is defined by
$$
f(x)=-\frac{1}{16}\sum\limits_{i=1}^5\nu_i^2\frac{S'(\mu_i)}{x-\mu_i}+k_1
x+k_0,
$$
where $k_1, k_0$ - some constants.

Calculation for a function (\ref{Z5roots}) gives
\be\label{siY} \sigma(x,y,\xi)= -\frac{1}{2}\sum\limits_{i=1}^5
\nu_i\log\frac{\sqrt{x-\xi}\sqrt{y-\mu_i}+\sqrt{y-\xi}\sqrt{x-\mu_i}}{\sqrt{x-y}\sqrt{\mu_i-\xi}},
\ee
$$
Y=\frac{1}{4}\sum\limits_{i=1}^N
\nu_i\frac{\sqrt{(x-\mu_i)(y-\mu_i)}}{(\xi-\mu_i)\sqrt{(x-\xi)(y-\xi)}}.
$$
Algebraic curve has the form of hyperelliptic curve of genus = 2
$$
\phi(Y,\xi)=S(\xi)Y^2+f(\xi)-\xi e_1+e_2=0
$$

Steklov top on $so(4)$ \cite{stek} is a particular case of Theorem 3.

\subsection{Class 2}
Functions $Z$ for the models of this class are the special cases of the functions  $Z$ of Class 1. 
But this Class contains much more parameters them Theorem 3.

Such functions $Z$ can be defined as the solutions of system
\be\label{constr}
Z_{x y}=\frac{Z_{x}-Z_{y}}{2(y-x)}=\frac{1}{3}
U(Z)\, Z_{x} Z_{y},
\ee
where $U$ - some functions of one variable.

{\bf Remark.} It easy to see that this class of solutions of Euler - Darboux equation $Z_{x y}=\frac{Z_{x}-Z_{y}}{2(y-x)}$ 
coincide with the class of solutions of the form
$$
Z=F\left(\frac{h(x)-h(y)}{x-y}  \right),
$$
where $F$ and $h$ - some functions of one variable and  $U=F''/F'^{2}$.

 {\bf Lemma.} {\it The system} (\ref{constr}) {\it is compatible if and only if}
$$
U=\frac{3}{2} \frac{B'}{B}, \qquad B(Z)=b_{2}Z^{2}+b_{1} Z+b_{0},
$$
{\it where} $b_i$ - {\it some constants}.

In a case deg $B=2$ 
 \be\label{casea}
Z(x,y)=\sqrt{(x-\mu_1)(y-\mu_1)}+\sqrt{(x-\mu_2)(y-\mu_2)}, \ee
where $  b_{2}=1, \quad b_{1}=0, \quad
b_{0}=-(\mu_{1}-\mu_{2})^{2}. $

If deg $B=1$, then \be\label{caseb}
Z(x,y)=\sqrt{x\,y}+\frac{1}{2}(x+y), \ee $b_{1}=1,$ $b_2=b_{0}=0$.

If deg $B=0,$ then
 \be\label{casec} Z(x,y)=x+y. \ee

{\bf 1.} Consider function $Z$ of the form (\ref{casea}). Then
$$
S(x)=(x-\mu_1)(x-\mu_2)P(x)+(x-\mu_1)^{3/2}(x-\mu_2)^{3/2} Q(x),
\qquad  \hbox{deg} P\le 3, \,\, \hbox{deg} Q\le 2,
$$
ш
$$
\begin{array}{c}
\displaystyle f(x)=f_0+f_1 x+k_2 (x-\mu_1)^{1/2}(x-\mu_2)^{1/2} +
\frac{(\mu_2-\mu_1) }{16}\, \Big\{
\frac{P(\mu_1)}{x-\mu_1}-\frac{P(\mu_2)}{x-\mu_2}\Big\} \, \\[5mm]
\displaystyle +\frac{(\mu_2-\mu_1)
}{32}\,(x-\mu_1)^{1/2}(x-\mu_2)^{1/2} \, \Big\{
\frac{Q(\mu_1)}{x-\mu_1}-\frac{Q(\mu_2)}{x-\mu_2}\Big\}.
\end{array}
$$
In a case when $Q=0,\, k_2=0,$  These formulas coinside with corresponding
formulas of Class 1. The functions $\sigma,\; Y$
are defined the same formula (\ref{siY}) as for Class 1 :
$$ \sigma(x,y,\xi)=
-\frac{1}{2}\sum\limits_{i=1}^2
\log\frac{\sqrt{x-\xi}\sqrt{y-\mu_i}+\sqrt{y-\xi}\sqrt{x-\mu_i}}{\sqrt{x-y}\sqrt{\mu_i-\xi}},
\quad Y=\frac{1}{4}\sum\limits_{i=1}^2
\frac{\sqrt{(x-\mu_i)(y-\mu_i)}}{(\xi-\mu_i)\sqrt{(x-\xi)(y-\xi)}}.
$$
Algebraic curve in this case has the form
\be\label{unicur1} \lbrack S_R(\xi)+\eta S_I(\xi)\rbrack Y^2-\lbrack
k_R(\xi)+\eta k_I(\xi)\rbrack=0, \ee
where
$$S_R(x)=(x-\mu_1)(x-\mu_2)P(x),\qquad   S_I(x)=(x-\mu_1)(x-\mu_2)Q(x),$$
$$k_R(x)=-e_2+e_1 x-f_0-f_1
x- \frac{(\mu_2-\mu_1) }{16}\, \Big\{
\frac{P(\mu_1)}{x-\mu_1}-\frac{P(\mu_2)}{x-\mu_2}\Big\},$$
$$k_I(x)=k_2-\frac{1}{32}(\mu_1-\mu_2)^2-\frac{1}{16}(\mu_1-\mu_2)\Big\{
\frac{Q(\mu_1)}{x-\mu_1}-\frac{Q(\mu_2)}{x-\mu_2}\Big\},$$
$$\frac{1}{\eta}=\frac{1}{\sqrt{\xi-\mu_1}\sqrt{\xi-\mu_2}}
\sqrt{1-\frac{(\mu_1-\mu_2)^2}{16(\xi-\mu_1)^2(\xi-\mu_2)^2Y^2}}.$$

Expressing $Y$ as a function of $(\xi,\eta)$ and substituting to
(\ref{unicur1}), we obtain 10-parameter cubic in $(\xi,\eta),$ variables.

So in a general case the curve $\phi(Y,\xi)=0$, is a covering over an elliptic curve
We obtain
$$ \eta=\frac{\xi-\mu_1}{\frac{\sqrt{x-\mu_1}}{\sqrt{x-\mu_2}}+\frac{\sqrt{y-\mu_1}}{\sqrt{y-\mu_2}}}+
\frac{\xi-\mu_2}{\frac{\sqrt{x-\mu_2}}{\sqrt{x-\mu_1}}+\frac{\sqrt{y-\mu_2}}{\sqrt{y-\mu_1}}},$$
therefore points $(\xi_1,\eta_1),(\xi_2,\eta_2),(\xi_3,\eta_3)$ lie on a straight line.

{\bf 2.} For the function  $Z$ of the form (\ref{caseb}) we have
$$
S(x)= x P(x)+x^{3/2}Q(x), \qquad  \hbox{deg} P\le 3, \,\,
\hbox{deg} Q\le 2,
$$
$$
f(x)=-\frac{1}{16x}P(x)-\frac{1}{32\sqrt{x}}Q(x)+f_1x+f_q\sqrt{x}+f_0.
$$
The function  $Y$ is defined by $\dsize
Y=\frac{\xi+\sqrt{x}\sqrt{y}}{4\xi\sqrt{x-\xi}\sqrt{y-\xi}}.$ 
The curve in this case can be written in the form (\ref{unicur1}), where
$$S_R(x)=xP(x),\qquad S_I(x)=x Q(x),$$
$$k_R(x)=-e_2+e_1 x-f_0-f_1x
+\frac{1}{16x}P(x),\qquad k_I(x)=\frac{1}{16x}Q(x)-f_q,$$
$$\eta=\frac{4Y\xi^{3/2}}{\sqrt{16Y^2\xi^2-1}}.$$
In $(\xi,\eta)$ variables it also has the form of arbitrary cubic.
Formula $\eta=\frac{\xi+\sqrt{xy}}{\sqrt{x}+\sqrt{y}}$ gives the fact that
points $(\xi_1,\eta_1),(\xi_2,\eta_2),(\xi_3,\eta_3)$ lie on a straight line

{\bf 3.} For the function $Z,$ given by (\ref{casec}), we obtain
$$
S(x)=s_6x^6+s_5x^5+s_4x^4+s_3x^3+s_2x^2+s_1x+s_0,
$$
$$
f(x)=-\frac{1}{40}S''(x)-\frac{1}{32\sqrt{x}}Q(x)+f_2x^2+f_1x+f_0.
$$
In this case $\dsize Y=\frac{1}{2\sqrt{x-\xi}\sqrt{y-\xi}}.$ Algebraic curve $$
S(\xi)Y^6-F(\xi)Y^4-\Big(\frac{1}{8}F''(\xi)+\frac{7}{1920}S^{IV}(\xi)-\frac{k_2}{2}\Big)\,Y^2-\frac{s_6}{64}=0,
\quad F(\xi)=-e_2+e_1\xi-f(\xi) $$ and in $(\xi,\eta),$  variables where
$\dsize \eta=\xi^2-\frac{1}{4Y^2},$ has the form of arbitrary cubic.
Because $\eta=\xi(x+y)-xy,$ points
$(\xi_1,\eta_1),(\xi_2,\eta_2),(\xi_3,\eta_3)$ belong to a straight line.

\subsection{Class 3}
 We introduce 'non-symmetrical Hamiltonian
(\ref{HKG})-(\ref{usl2}) if
$\,\, S_{1}(x)\ne S_{2}(x), \quad \hbox{or} \quad f_{1}(x)\ne
f_{2}(x)$.

{\bf Theorem 4.} \cite{marsok2} {\it In non-symmetrical case the functions
} $Z$, $S_{i},$ $f_{i}$ {\it is the solutions of } (\ref{usl1}),
(\ref{usl2}) {\it if and only if}
$$
\delta=0, \qquad g=\frac{1}{H}, \qquad S_{1,2}=W\,H \pm M H^{3/2},
\qquad f_{1,2}= -\frac{4 W}{H}\mp 2 M H^{-1/2}\pm a H^{1/2},
$$
{\it where} $g$ - {\it generation function of} $Z$,
$$
W(x)=w_{3} x^{3}+w_{2} x^{2}+w_{1} x+w_{0}, \qquad H(x)=h_2x^2+h_1x+h_0,
$$
$$
M(x)=m_{2} x^{2}+m_{1} x+m_{0}.
$$
{\it Here} $w_{i},h_{i},m_{i},a$ - {\it some constants}.

Consider the general case   $H(x)=(x-\mu_1)(x-\mu_2).$
Algebraic curve in this case is defined by 
\be
\begin{array}{c} \label{cur_non} \dsize \Psi(\xi,
Y)=-e_2+e_1\xi-\frac{R\,W(\xi)}{2(\xi-\mu_1)(\xi-\mu_2)(\mu_2-\mu_1)}+ \\[7mm]
\dsize 4
M(\xi)\sqrt{2}Y\frac{\sqrt{\xi-\mu_1}\sqrt{\xi-\mu_2}}{(\mu_2-\mu_1)^{3/2}}\sqrt{R}+
8b\sqrt{2}Y\frac{(\xi-\mu_1)^{3/2}(\xi-\mu_2)^{3/2}}{\sqrt{R}\sqrt{\mu_2-\mu_1}}=0,\end{array}
\ee where
$$Y=\frac{\sqrt{(x-\mu_1)(y-\mu_1)}}{(\xi-\mu_1)\sqrt{(x-\xi)(y-\xi)}}-\frac{\sqrt{(x-\mu_2)(y-\mu_2)}}{(\xi-\mu_2)\sqrt{(x-\xi)(y-\xi)}}
,\quad R=16(\xi-\mu_1)^2(\xi-\mu_2)^2Y^2-(\mu_1-\mu_2)^2.$$
Substituting
$$
Y=\frac{1}{4}\frac{(\mu_1-\mu_2)^{\frac{3}{2}}\eta}
{(\xi-\mu_2)(\xi-\mu_1)\sqrt{\eta^2(\mu_2-\mu_1)-8(\xi-\mu_1)(\xi-\mu_2)}}
$$
into (\ref{cur_non}), We obtain the cubic in variables $(\xi,\eta)$ with a full set
of ten independent parameters.
It easy to see that  $\eta=a(x,y)\xi+b(x,y),$ where $a,b$ - some functions.

Therefore in the cases of Class 2 and 3 the algebraic curve is
non-hyperelliptic covering over the elliptic curve. The dynamics of the points
$(\xi_1,Y_1),(\xi_2,Y_2),(\xi_3,Y_3)$ on this curve (see. theorem 2)
defines the following condition: their projection on the elliptic base
$(\xi_1,\eta_1),(\xi_2,\eta_2),(\xi_3,\eta_3)$ lies on the straight line.

{\bf Hypothesis 2.} Any pair of the Hamiltonians
(\ref{HKG})-(\ref{usl2}) belongs to one of above three classes.

\section{Appendix 1.    Steklov top}\label{stcase}

We show that the case of Steklov top on $so(4)$ is a particular case of Class 1
after restriction on the symplectic leafs.
Hamiltonian and the additional integral in this case have the form
$$
H=({\bf {\bf \vec{S}}}_1,A{\bf \vec{S}}_1)+({\bf \vec{S}}_1,B{\bf
\vec{S}}_2),\qquad K=({\bf \vec{S}}_1,\bar{A}{\bf \vec{S}}_1)+({\bf
\vec{S}}_1,\bar{B}{\bf \vec{S}}_2),
$$
where
$$A=-\alpha^2\,diag(\frac{1}{\alpha_1^2},\frac{1}{\alpha_2^2},\frac{1}{\alpha_3^2})
,\qquad  B=\alpha\, diag(\alpha_1,\alpha_2,\alpha_3), $$$$
\bar{A}=-diag(\alpha_1^2,\alpha_2^2,\alpha_3^2),\qquad
\bar{B}=\alpha\,diag(\frac{1}{\alpha_1},\frac{1}{\alpha_2},\frac{1}{\alpha_3}),$$
$\alpha=\alpha_1\alpha_2\alpha_3.$ Here ${\bf {\bf \vec{S}}}_i$ -
three-dimensional vectors with components $S_i^{\alpha}.$ It is easy to see that,
$H$ and $K$ commute under a spin Poisson bracket
$$
 \lbrace S_i^{\alpha},S_j^{\beta}\rbrace=
\kappa\,\varepsilon_{\alpha\beta\gamma}\delta_{ij}\, S_i^{\gamma}.
$$
It is convenient to chose $\kappa=-2 i.$

We fix the Casimirs for spin bracket: ${\bf
(\vec{S}}_k, {\bf \vec{S}}_k)=j_k^2, \,\, k=1,2.$ Then the formulas
$$ {\bf
\vec{S}}_k=\pi_k{\bf \vec{K}}(Q_k)+\frac{j_k}{2}{\bf
\vec{K}}'(Q_k),\qquad \hbox{\rm уфх} \qquad {\bf
\vec{K}}(Q)=((Q^2-1),\,i(Q^2+1),\,2Q),
$$
define the Darboux coordinate  $\pi_{1},\pi_{2}, Q_{1},Q_{2}$  for the simplectic leaf of Poisson manifold with coordinates${\bf
\vec{S}}_k, \,\, k=1,2.$ As this transformation is linear by momenta $\pi_k$,  as a result, we obtain
a pair of commuting Hamiltonians quadratic in momenta under the bracket $\lbrace
\pi_{\alpha},Q_{\beta}\rbrace=\delta_{\alpha \beta}$. Consider the canonical transformation
that transforms their pair to the form (\ref{HKG})-(\ref{tails}), (\ref{sym}).

We apply the canonical transformation
$$
P_1=\pi_1\sqrt{r(Q_1)},\quad P_2=\pi_2\sqrt{R(Q_2)},\qquad
dX=\frac{dQ_1}{\sqrt{r(Q_1)}}, \quad dY=\frac{dQ_2}{\sqrt{R(Q_2)}},
$$
where
$$
r(Q_1)=({\bf \vec{K}}(Q_1),A{\bf \vec{K}}(Q_1)),\qquad R(Q_2)=({\bf
\vec{K}}(Q_2),\bar{A}{\bf \vec{K}}(Q_2)),
$$
to obtain
$$
H=P_1^2+2P_1P_2V+j_2P_1V_Y+j_1P_2V_X+\frac{1}{2}j_1j_2V_{X,Y}+
\frac{j_1^2}{6}\Big(\frac{g_1''(X)}{g_1(X)}-\frac{3}{2}(\frac{g_1'(X)}{g_1(X)})^2\Big),
$$
$$
K=P_2^2+2P_1P_2W+j_2P_1W_Y+j_1P_2W_X+\frac{1}{2}j_1j_2W_{X,Y}+
\frac{j_2^2}{6}\Big(\frac{g_2''(Y)}{g_2(Y)}-\frac{3}{2}(\frac{g_2'(Y)}{g_2(Y)})^2\Big).
$$
where
$$
V(X,Y)=\frac{({\bf \vec{K}}(Q_1),B{\bf
\vec{K}}(Q_2))}{\sqrt{r(q_1)}\sqrt{R(Q_2)}},\qquad
W(X,Y)=\frac{({\bf \vec{K}}(Q_1),\bar{B}{\bf
\vec{K}}(Q_2))}{\sqrt{r(Q_1)}\sqrt{R(Q_2)}},
$$
$g_1(X)=\sqrt{r(Q_1)},\;g_2(Y)=\sqrt{R(Q_2)}.$

We apply the canonical transformation 
$(P_1,P_2,X,Y)\rightarrow (p_1,p_2,x,y)$ of the form
$$
dX=\frac{1}{2}\Big(\frac{dx}{\sqrt{S(x)}}+\frac{dy}{\sqrt{yS(y)}}\Big),\qquad
dY=-\frac{1}{2}\Big(\frac{dx}{\sqrt{xS(x)}}+\frac{dy}{\sqrt{S(y)}}\Big),
$$
$$
P_1=\frac{2}{\sqrt{x}-\sqrt{y}}\Big[ \Big(p_1-\frac{j_1+j_2}{4(x-y)}
\sqrt{\frac{y}{x}}\Big)\sqrt{S(x)} -\Big(p_2+\frac{j_1+j_2}{4(x-y)}
\sqrt{\frac{x}{y}}\Big)\sqrt{S(y)}\Big],
$$
$$
P_2=\frac{2}{\sqrt{x}-\sqrt{y}}\Big[ \Big(p_1-\frac{j_1+j_2}{4(x-y)}
\sqrt{\frac{y}{x}}\Big)\sqrt{y\,S(x)}
-\Big(p_2+\frac{j_1+j_2}{4(x-y)}
\sqrt{\frac{x}{y}}\Big)\sqrt{x\,S(y)}\Big],
$$
to obtain (\ref{HKG})-(\ref{tails}), (\ref{sym}), where $$ S(x)=-4\,
x(1+\alpha_1^2x)(1+\alpha_2^2x)(1+\alpha_3^2x),\qquad
Z(x,y)=-\frac{1}{2}j_1(x+y)-j_2\sqrt{xy}, $$
$$
f(x)=\frac{1}{4}\Big(j_1^2\alpha^2x^2+\frac{j_2^2}{x}\Big)
-j_2^2\frac{1}{4}\alpha^2\Big(\frac{1}{\alpha_1^2}+\frac{1}{\alpha_2^2}+\frac{1}{\alpha_3^2}\Big)\,x.
$$

{\bf Acknowledgements.} The author are grateful to E V Ferapontov for useful discussion.
The research was partially supported by RFBR grant 05-01-00189 and NSh 6358.2006.2.

\end{document}